\documentclass[preprint,12pt]{elsarticle}

\usepackage[centertags]{amsmath}
\usepackage{amssymb}
\usepackage{epsfig}
\usepackage{color} 
\usepackage{epstopdf}
\usepackage{enumerate}
\usepackage{enumitem}
\usepackage{rotating}
\usepackage[T1]{fontenc}
\usepackage[utf8]{inputenc}
\usepackage{url}
\usepackage{hyperref}
\usepackage{lineno}

\journal{Applied Energy}

\begin{document}
	
\begin{frontmatter}



\title{A Genetic Algorithm Approach for Modelling Low Voltage Network Demands}


\author[label1]{Georgios~Giasemidis}
\ead{georgios@countinglab.co.uk}
\author[label2]{Stephen~Haben}
\ead{Stephen.Haben@maths.ox.ac.uk}
\author[label2]{Tamsin~Lee}
\ead{Tamsin.Lee@maths.ox.ac.uk}
\author[label1]{Colin~Singleton}
\ead{colin@countinglab.co.uk} 
\author[label2]{Peter~Grindrod}
\ead{grindrod@maths.ac.uk}

\address[label1]{CountingLab LTD, Reading, UK}
\address[label2]{Mathematical Institute, University of Oxford, Oxford, UK}

\begin{abstract}
Distribution network operators (DNOs) are increasingly concerned about the impact of low carbon technologies on the low voltage (LV) networks. More advanced metering infrastructures provide numerous opportunities for more accurate load flow analysis of the LV networks. However, such data may not be readily available for DNOs and in any case is likely to be expensive. 
Modelling tools are required which can provide realistic, yet accurate, load profiles as input for a network modelling tool, without needing access to large amounts of monitored customer data. 
In this paper we outline some simple methods for accurately modelling a large number of unmonitored residential customers at the LV level. We do this by a process we call \emph{buddying}, which models unmonitored customers by assigning them load profiles from a limited sample of monitored customers who have smart meters. Hence the presented method requires access to only a relatively small amount of domestic customers' data. The method is efficiently optimised using a genetic algorithm to minimise a weighted cost function between matching the substation data and the individual mean daily demands. Hence we can show the effectiveness of substation monitoring in LV network modelling. Using real LV network modelling, we show that our methods perform significantly better than a comparative Monte Carlo approach, and provide a description of the peak demand behaviour.
\end{abstract}


\end{frontmatter}

\section{Introduction}
\label{sec:introduction}

Distribution network operators (DNOs) are increasingly interested in modelling the demand behaviour at the low voltage (LV) level to improve network planning and management. In the next few decades, electricity demand is expected to increase and become more irregular with the uptake of low carbon technologies such as electric vehicles and photovoltaics \cite{EATech}. The grids' changing demands mean DNOs require accurate network models and load flow tools to evaluate the stability and remaining headroom on their LV networks. As well as load flow analysis, such models could be implemented to validate potential energy reduction schemes such as demand side response or energy storage devices \cite{Lyons2015}.

Up until recently, DNOs have only had access to half hourly energy data from heavy commercial customers, but for residential customers, DNOs tend to only have quarterly cumulative readings, which are often estimates. However, with the roll-out of smart meters and other advanced metering infrastructures (such as LV substation monitors \cite{Li2015a}) such information could be utilized by a DNO to help understand LV network headroom \cite{Harrison2005}, the effect of energy storage \cite{Rowe2014}, and the network renewable capacity \cite{Ochoa2010}. For example, by considering fully monitored customers on an LV network, the authors in \cite{Hoogsteen2015} and \cite{Hoogsteen2013} were able to test network modelling tools and perform load flow analysis on LV networks of 83 and 121 households respectively. Unfortunately, available individual household data is likely to be expensive~\cite{EATech} for a DNO which is only likely to have access to high resolution substation monitoring or aggregates of the smart meter data. 

With limited smart meter data, customer demand must be simulated or modelled. Residential demand is naturally more volatile than higher voltage, aggregated demand but it is exactly these realistic features (such as the peaks and troughs in demand) which are necessary for inclusion in an LV network modelling tool. Historically, in the UK, DNOs have estimated LV network demands through the after diversity maximum demand (ADMD) procedure or modelled aggregate profiles using the specifications of ACE Report No. 49 (ACE49) for the design of LV radial distribution networks \cite{Richardson2010, ENA1981}. These procedures can give estimates for the maximum demands at the aggregate level for residential networks of various sizes and, in the case of ACE49, an estimate of standard daily profiles for a central winter period. However the changing nature of energy demand was not anticipated when these tools were developed and hence they may not be suitable for future network design in a low carbon economy. For these reasons advanced modelling tools are required for simulating customers' demands and generation on the LV networks.

A common approach given in the literature is to create and utilise standardised, or ``typical'' profiles for customers \cite{Swan2009}.  For example, in the UK, electricity suppliers assume two high-level Elexon groupings for domestic customers; Standard and Economy $7$ \cite{Elexon2013,Dent2011b}. Much recent research has focused on creating typical household profiles by clustering customers' demand based on smart meter data \cite{Flath2012, Stephen2014, Shengyan2011}, also see references within \cite{Frame2016}. A variety of techniques and methods have been employed to create such clusters including Gaussian Processes \cite{McLoughlin2013}, Gaussian Mixture models \cite{Haben2016}, k-means and self-organising maps \cite{Rasanen2010}, k-mediods \cite{McLoughlin2015} and principle component analysis \cite{Abreu2012}. For use within a network modelling environment, the aim is to link such clusters to other available characteristics that would potentially be available to a DNO such as socio-demographics or household properties. However, energy behavioural clusters have shown only a weak correlation with socio-demographic groups \cite{Haben2013}, tariff types \cite{Haben2016} and even mean daily demands \cite{McLoughlin2012}. In general it has been shown that variability in residential demand has been difficult to classify, despite similar household size, numbers of occupants and dwelling type \cite{Morley2011}. Hence it is far from straightforward to assign such profiles to unmonitored customers to use within an LV network modelling tool. Further, such standardised profiles smooth important features such as demand peaks and troughs and thus do not take into account the natural volatility of residential demand. This limits their effectiveness in LV network modelling tools. 

A final approach is to utilise limited amounts of monitoring to simulate network demand and generation for all unmonitored customers on the network.
One possible way is to use a bottom-up approach which creates household level demand through the aggregation of the demand profiles of individual appliances \cite{Capasso1994, Paatero2006, Richardson2010, Widen20101880, Muratori2013465}. These approaches model household demand from (i) the set of appliances in the household (ii) the electricity demand of these appliances (iii) the use of the appliances. They require sample datasets of a household's major appliances and behavioural patterns. The latter often come from survey data, which could potentially be costly to regularly update. Therefore, such models result in (i) greater model complexity and (ii) input data requirements that are greater than top-down models \cite{Muratori2013465}.
At the other extreme, the authors in \cite{Li2015a} and \cite{Rigoni2016} have analysed monitoring at the low voltage level to find typical substation level profiles and link them to the types of connected customers (residential, commercial, industrial, etc.). Although these methods do not model individual customers they could potentially be combined with other techniques such as those described here, for improving modelling at the aggregate level. Another very common approach, which uses limited monitoring data, is a Monte Carlo simulation that generates several thousand random assignments from a sample set of smart meter data to understand the range of LV network impacts \cite{Neaimeh2015, Navarro2014, Navarro2013}. For example, the authors in \cite{Neaimeh2015} used this approach on two LV networks of 189 and 288 customers respectively to understand the impact of large electric vehicle (EV) uptakes. An advantage of this method is that the random nature of demand is taken into account and is highly versatile, e.g. it can be combined with the EV or any other low carbon technology data. A disadvantage is that several thousand implementations of the load flow analysis could be computationally expensive, especially if the DNO is planning to simultaneously analyse thousands of LV networks. Secondly, network impacts could be exaggerated since individual characteristics of the customers are not taken into account. 

In this study we present a new method for modelling unmonitored customers on LV networks that assigns (``buddies'') profiles of monitored customers to unmonitored households utilising substation monitoring and limited customers' information, such as their quarterly meter readings. The solution is found by optimising the match between the half-hourly demand at the aggregate level and approximations to the mean daily demand of customers. The optimisation is implemented through a genetic algorithm. Hence we can populate a network modelling environment using either information readily available to a DNO (quarterly meter
readings) or data that is, or soon will be, available such as substation or aggregate smart meter data.

This paper makes a number of novel contributions to the modelling of LV network. Firstly, we present a new framework, called buddying, for modelling both LV substations' and individual households' electricity demand. The method provides real residential demand profiles from a diverse sample of monitored customers, which are essential for load flow analysis, using minimal monitoring. This distinguishes the method from clustering approaches which use smooth (hence less realistic) profiles. 
We show that the results are significantly more accurate than a comparable Monte Carlo approach at the aggregate (LV) level \cite{Neaimeh2015, Navarro2014, Navarro2013}. In fact, genetic algorithms ensure faster convergent to the optimal solution than the Monte Carlo methods. In particular we show how the Monte Carlo approach could potentially result in less accurate aggregated (feeder) profiles and hence misleading results from load flow analysis.
Secondly, to the best of our knowledge, our study is the first that considers an extended network consisting of hundreds of feeders and is not restricted to a few representative substations. In particular, we show that the method is robust across a large range of different types of LV feeders, which allows us to draw statistically significant conclusions and understand the effectiveness of modelling at the LV level as a function of the number of customers. This can also help DNOs make informed management and planning decisions for different types of networks. Finally, we show the importance of correct connectivity information of the LV network to the accuracy of the modelling. For most networks connectivity at the phase level is unknown so we show how this affects the overall accuracy of our method. 

This paper is organised as follows. In Section \ref{sec:Methods} we introduce the two main methods developed in this study, present the corresponding algorithms and discuss the evaluation scores. Section \ref{sec:Results} presents the results of our modelling on real data, whereas in Section \ref{sec:Pseudo_feeders} we discuss the analysis on pseudo-feeders which serve as toy-models for evaluating the accuracy on individual profiles. Finally, we summarise and identify future work in Section \ref{sec:Discussion}.

\section{Methodology}
\label{sec:Methods}

Suppose we have $M$ residential customers on a low voltage feeder (or phase), labelled $c_j$, $j=1, \ldots, M$, for which we have their mean daily demand usage, $U_j$ (in kWh), estimated from their quarterly meter readings. Further, we assume access to half hourly energy data for $N$ residential customers for $d$ days (at least a year). Let 
\begin{equation} 
\mathcal{P}=\{\mathbf{p}_i=\left (p_i(1), \ldots , p_i(48d)\right)^T  \in \mathbb{R}^{48d}, i=1, \ldots , N\}
\label{eq:monitoredset}
\end{equation}
be the known set of profiles of energy data for $N$ monitored customers. The location and connectivity of these customers are irrelevant to the model. The mean daily demand of the monitored customers, $\hat{U}_j$, is estimated from the half-hourly data, i.e.
\begin{equation}
\hat{U}_j:=\frac{1}{d}\sum_{k=1}^{48d}p_i(k).
\end{equation}
If a customer $c_j$ on the current feeder is monitored, then we set its mean daily demand to be $U_j = \hat{U}_j$. Finally, one of the methods requires substation monitoring data at half hourly resolution for the same time period denoted as 
\begin{equation}
\mathbf{s}=\left(s(1), \ldots, s(H)\right)^T
\end{equation}

The methods proposed in this study assigns a profile $\mathbf{p}_i \in \mathcal{P}$ to every unmonitored customer $c_j$ based on minimal customer's information, e.g. quarterly meter readings. The half-hourly profiles, $\mathbf{p}_i$, populate the network of unmonitored customers, $c_j$, and ensure that each customer is given a real profile.

We split all customers, both monitored and unmonitored, into seven ``customer groups'' defined by the customer's council tax band and Elexon profile class\footnote{There are eight generic Elexon profile classes representative of large populations of similar customers. Two classes correspond to domestic customers and distinguish between two tariffs, ``Standard'' and ``Economy 7''. The latter provides cheaper rates overnight at the expense of increased day-time charges.}
\cite{Elexon2013,Dent2011b}. The grouping ensures similar property types are buddied and reduces the computational cost of the optimisation, see \ref{sec:Grouping} for further details. The same monitored customer can be buddied to multiple unmonitored customers. 

To buddy, the feeder (or phase) connectivity must be known. This is generally accurate at the feeder level but at the phase level this is less certain. For these reasons we focus our results at the feeder level and the effect of buddying at the phase level is discussed in Section \ref{comparison_to_phaselevel}.

In the current study we do not consider non-domestic properties and focus on feeders with residential properties only, due to the lack of monitored commercial customers in the dataset. Although there are techniques to estimate the profile of commercial customers, such as~\cite{Lee2014}, they introduce further ambiguities in the evaluation of our methods. Improved commercial demand modelling will be a topic of future work.

\subsection{Method 1: the simple buddy}

The first buddying method, which we refer to as the simple algorithm (SA), uses only the mean daily demand $U_j, j = 1 \ldots, M$ for each customer on a feeder or phase. In this scheme, an unmonitored customer, say $c_j$ which is a member of customer group $g$ (see \ref{sec:Grouping}), is assigned with the profile $\mathbf{p}_i \in \mathcal{P}$ such that
 \begin{equation}
 i=\underset{k\in I_g}{\operatorname{argmin}}  |U_j-\hat{U}_k|,
 \label{eq:simple_buddying}
 \end{equation} 
where $I_g$ is the index for profiles in group $g$. Hence, the buddy is the monitored customer profile with the closest mean daily demand. The advantage of this method is that it only uses information readily available to a DNO. A major disadvantage is that household quarterly meter reads are often based on, possibly inaccurate, estimates. In addition, it is known that daily demand weakly correlates with intra-day demand \cite{McLoughlin2012}. 

\subsection{Method 2: a genetic algorithm optimised buddy} 
\label{GAbuddsec}

In this buddying we use half hourly energy data of an LV feeder to create a new buddying constrained by the customers' mean daily usage (as in the SA buddying) and the total half hourly demand on each feeder. We optimise our buddy using a genetic algorithm (GA), which  mimics the process of natural selection.

\begin{figure}
	\begin{center}
	\includegraphics[scale=0.4]{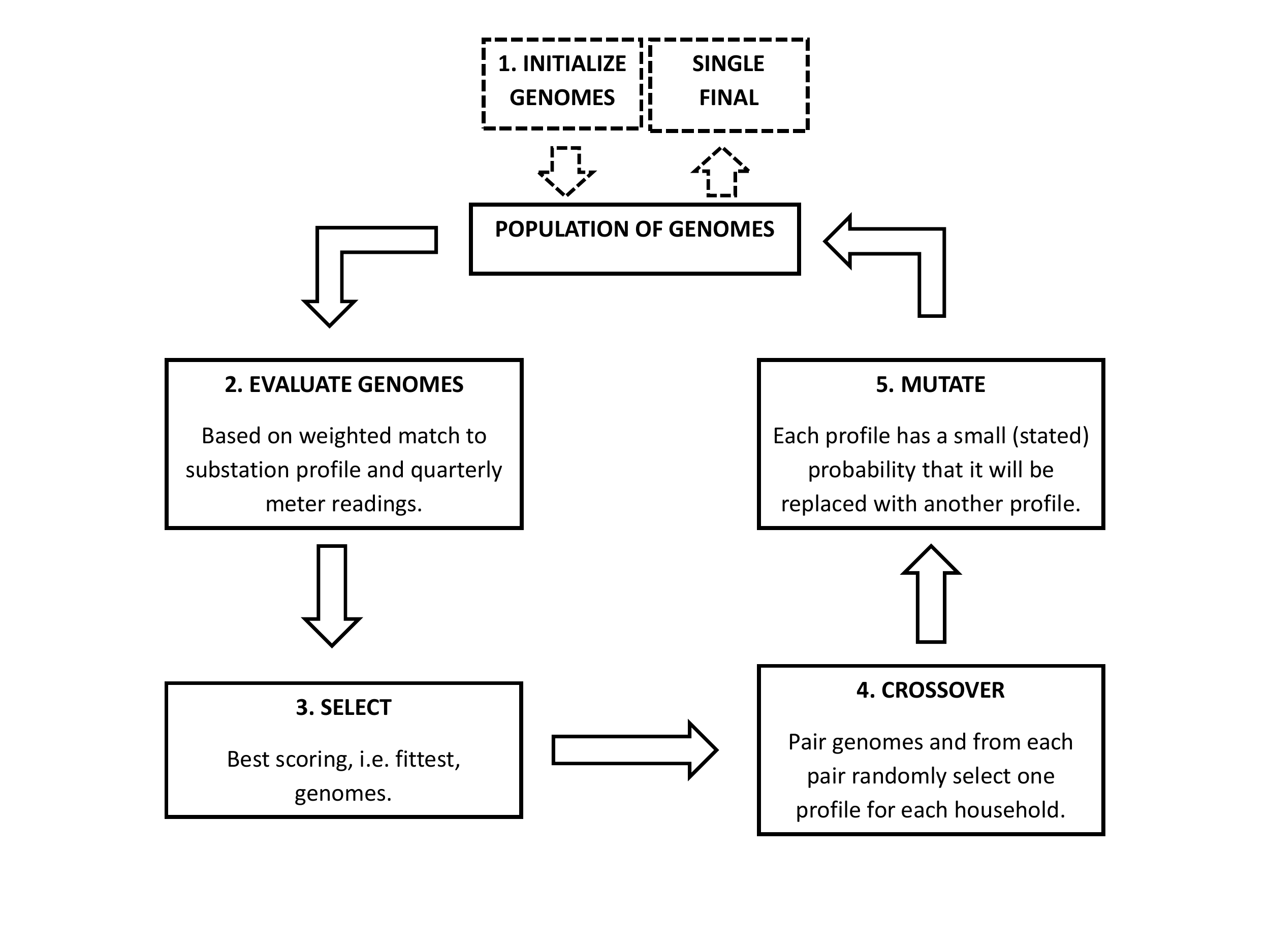}
	\caption{Flow chart of genetic algorithm optimization.}
	\label{fig:GAalgo}
	\end{center}
\end{figure}

A time period of $d_t$ days, totalling $H=48d_t$  half-hour readings, is chosen for training. The genetic algorithm proceeds by creating updates of several collections of monitored customers according to how well they score according to a ``fitness function''. Our fitness function is a weighted measure between how well the aggregation of the buddied profiles matches the substation actuals, and how closely the buddies match the daily average estimates of the unmonitored customers. For a set of buddied profiles $\mathcal{\hat{P}}=\{ \mathbf{p}_{k_1}, \ldots ,\mathbf{p}_{k_M} \} \subset \mathcal{P}, k_i \in \{1, \ldots, N\}$
assigned to customers $\mathcal{C} = \{c_1, \ldots, c_M\}$,  the fitness function is given by
\begin{equation}
F(\hat{\mathcal{P}}, \mathcal{C}, \mathbf{s})=(1-w)\sum_{t=1}^{H} \frac{\|a(t) - s(t)\|}{S} +w\sum_{j=1}^M \frac{\|U_j-\hat{U}_{k_j}\|}{D},
\label{eq:CostFctn}
\end{equation}
where $a(t) = \sum_{j=1}^{M}p_{k_j}(t)$ is the aggregated demand of the buddied profiles on the feeder at half hour $t$, $S=\sum_{t=1}^{H}s(t)$ and $D = \sum_{j=1}^MU_j$ are normalisation factors and $w \in [0,1]$ is a weighting parameter. 
When $w = 0$ the fitness function considers only matching to the feeder data, and the algorithm finds the optimal collection of buddies whose aggregated demand matches the feeder profile $\mathbf{s}$. At the other extreme, when $w = 1$, the first term vanishes and the algorithm converges to the selection of monitored customers with the minimum mean daily demand difference, i.e. the simple buddying method. The GA buddying method is outlined below and summarised in Figure \ref{fig:GAalgo}.

\begin{enumerate}[label=Step \arabic*:]
	\item Initialize the buddy. Create $G$ \emph{genomes} each consisting of $M$ randomly selected profiles from $\mathcal{P}$ for a training period of $H$ half hours  for each customer $c_j, j = 1, \ldots, M$. The selection of the buddies is only restricted so that the buddies belong to the same group as customer $c_j$.
	\item The fitness of each genome  is evaluated using the fitness function (\ref{eq:CostFctn}).
	\item Select the best-scoring (fittest) genomes.
	\item To create each of the $G$ next generation genomes, two of the current best $G^\prime < G$ genomes are randomly selected for \emph{crossover}. Common profiles are retained while the remaining profiles are selected randomly from one or the other genome.
	\item The new genome is \emph{mutated} by replacing each profile with a probability $p$ with a new profile (from the same group).
	\item Repeat steps 1 to 3 for 100 generations. 
\end{enumerate}

We are free to choose the probability of mutation, which is initially set to $p = 0.1$ and slowly decreases as the algorithm progresses. A mutation rate too low and the genomes may lose variation, too high and we may remove good solutions from the population. 
For step 3, after $40$ iterations we reset the genomes, whilst retaining the best genome, to reduce the chances of finding a local minimum. 

The method requires LV monitoring, which could be expensive, but in practice DNOs could target substations where buddying is most effective. An advantage of this method is the reduced requirement for smart meter data.

\subsection{Evaluating the results}
\label{Evalution}

There are two ways of measuring the accuracy; how closely we fit the substation profile and how closely we match the individual profiles. Since we only have a limited number of monitored customers on the monitored feeders we cannot assess the household level accuracy in a significant way and so we focus on the feeder level accuracy. Of course this means that the substation level (i.e. $w=0$ in eq. \eqref{eq:CostFctn}) buddy will be favoured. To address household level accuracy we consider some pseudo-feeders in Section \ref{sec:Pseudo_feeders}.
To assess the fit to the substation demand we consider a relative average absolute error between the actual substation demand and the aggregated buddied profiles. We define the relative mean absolute error (RMAE) as
\begin{equation}
RMAE = \frac{1}{H \, S}\sum_{t=1}^{H} \|a(t) - s(t) \|.
\label{eq:errormeasure1}
\end{equation}
Dividing by the mean feeder demand allows for relative comparison of substations with different demands. 

A DNO is particularly interested in the peak demand on the network for planning and managing purposes, especially when considering long term scenarios \cite{Hattam2017}. In this study we focus on the magnitude peak demand error by considering the relative peak demand error (RPDE),
\begin{equation}
RPDE=\frac{\max_{t=1, \ldots, H} s(t) - \max_{t=1, \ldots, H}a(t)}{\max_{t=1, \ldots, H}s(t)}.
\label{eq:errormeasure2}
\end{equation}

For our purposes, the timing of the peak is of secondary interest in a load flow analysis tool operated by DNOs, in contrast to electricity price forecasting, in which case timing is of major importance \cite{Muratori2014546}. However, the method is quite versatile and we can increase emphasis on accurate peak model by using a $p>1$ in the $p$-norm of the fitness function \eqref{eq:CostFctn}. We leave a further and more detailed analysis on the timing of the peak error for future work.

\subsection{Data, calibration and validation periods}
\label{sec:Data}

Our dataset consists of half hourly energy data from $46$ LV substations (in Bracknell, UK), collected as part of the Thames Valley Vision project.\footnote{See \url{http://www.thamesvalleyvision.co.uk/}} In total, this is $122$ feeders and $366$ phases. The whole trial period we consider is from 20th March 2014 to 22nd September 2015 inclusive ($d = 551$ days). We also use $N = 242$ monitored domestic profiles at half-hourly resolution for the same trial period. We pre-processed the raw data to replace missing values, outliers and anomalous readings with the average load from similar hours.

The available data spans six seasons and we investigate how buddying varies with season and determine the best season for calibrating the buddying. For training, we choose the six season to commence on 24/03/2014, 23/06/2014, 29/09/2014, 05/01/2015, 04/05/2015 and 27/07/2015 respectively. 
We must also optimise the length of the training period (in weeks). We restrict the length of the training period to less than 9 weeks to avoid overlapping periods for different seasons and reduce seasonality effects during training.
We also explore how the weighting parameter affects buddying using a range of weighting parameters from $0$ to $1$ in increments of $0.1$. 
To summarise, we perform buddying for each combination of the six seasons, eight different lengths of the training period and eleven values of the weight parameter, resulting in $6\times8\times11 = 528$ models for each of the $122$ feeders.

The test period is chosen to be an entire year (to reduce biases due to seasonal effects) from 01/09/2014 until 31/08/2015.

\section{Results on real feeders} 
\label{sec:Results}

\begin{figure}
	\includegraphics[width=\columnwidth]{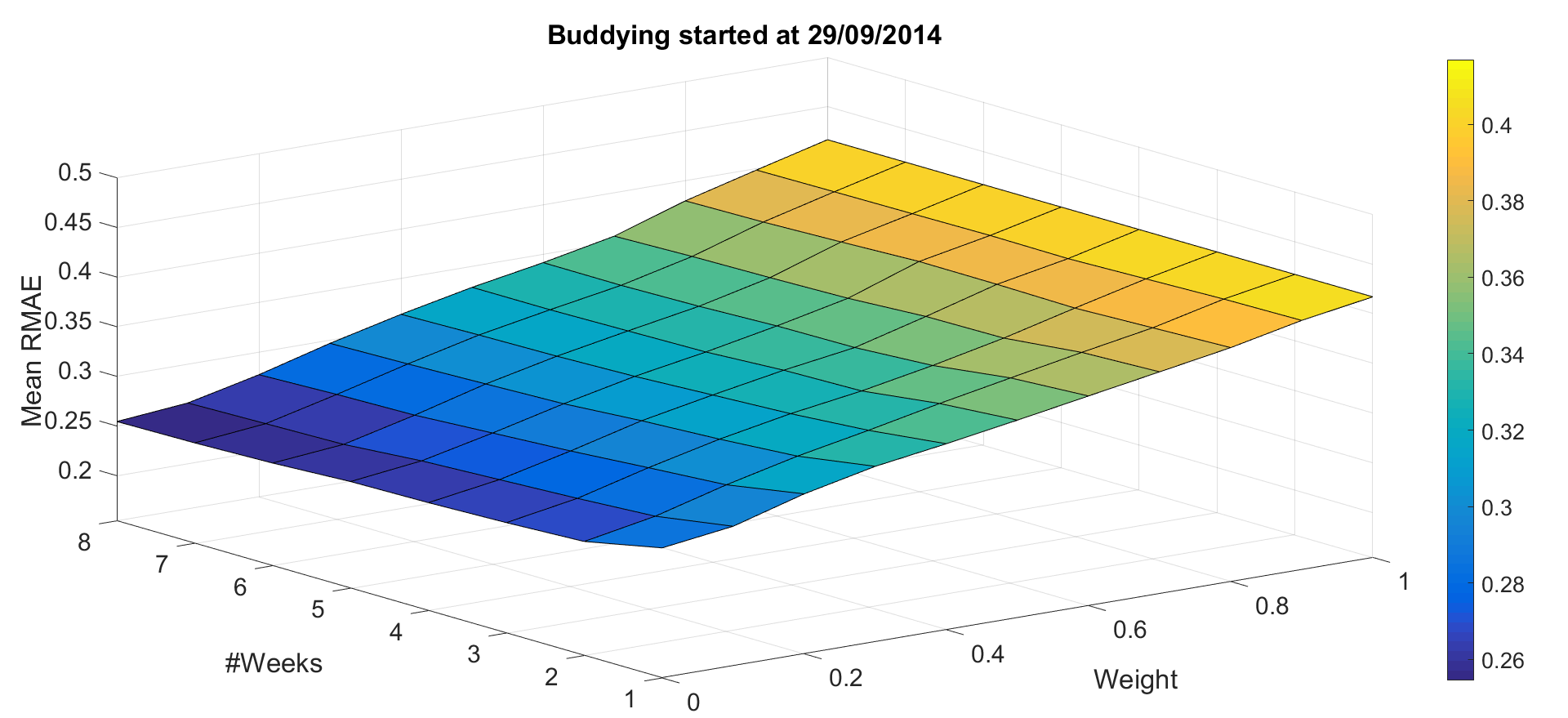}
	\caption{Average RMAE errors of all feeders for season commencing on 29/09/2014 (autumn 2014). The z-axis is the average error score value, the x-axis, labelled as ``Weight'', corresponds to the weight in the fitness function of the genetic algorithm. The y-axis, labelled as ``\#Weeks'', is the number of weeks used in the training of buddying, each starting at the first date of the corresponding season. Other seasons give similar shape of the error plot.}
	\label{fig:3Derrors}
\end{figure}

We  first consider the average feeder level RMAE error \eqref{eq:errormeasure1} across all feeders for different seasons, weights and lengths of training periods.  We plot how these errors vary with weight and the duration of training (in weeks) for the season starting on 29/09/2014 in Figure \ref{fig:3Derrors}. From this figure we observe a pattern (which is common in all seasons); the error decreases towards smaller weights (i.e. less emphasis on the mean daily usage) and longer periods of training. These observations are consistent with our expectations. First, the error score \eqref{eq:errormeasure1} favours the buddying with zero weight, i.e. training the buddying on feeder readings only. Second, the RMAE error \eqref{eq:errormeasure1} favours a longer training period too. 

From these six seasons, autumn 2014 (starting on 29/09/2014) and winter 2015 (starting on 05/01/2015) achieve the lowest and second-lowest RMAE scores respectively. One can argue that the cold periods in UK have the highest volatility,\footnote{Air conditioning is uncommon in UK during summer and there is no evidence for affecting the electricity load. This behaviour results in smoother profiles in summer than in winter.} therefore training the buddying during these periods ensures the fittest buddies with respect to the feeder total.

For the remainder of this section we further focus on the autumn 2014 season which delivers the most accurate results. Training the buddying for eight weeks, the simple buddy (i.e. unit weight) average error is 0.417, whereas for zero weight the average error reduces to  0.255, showing a decrease by 39\%. To conclude we argue that the GA buddying for zero weight, compared to the SA, improves the accuracy of buddying with respect to the feeder total by almost 40\%. 

In \cite{Sevlian2014}, the authors studied the effect of aggregation on one hour ahead load forecasting. They report that for an aggregate of roughly 20 households, the MAPE error of a short term load forecast is 12\%. In our case, the average feeder has 35 customers and the average MAPE for long-term (one year) buddying is 27\%. The error comparison should be treated with caution since we are comparing long term, year long, estimates (buddying) to short term, one hour ahead, forecasting. For further analysis of the errors we compare our results against a Monte Carlo approach in Section \ref{sec:benchmark}.

\subsection{Distribution of errors}
\label{sec:disterrors}
\begin{figure}
\begin{center}
	\includegraphics[width = \columnwidth]{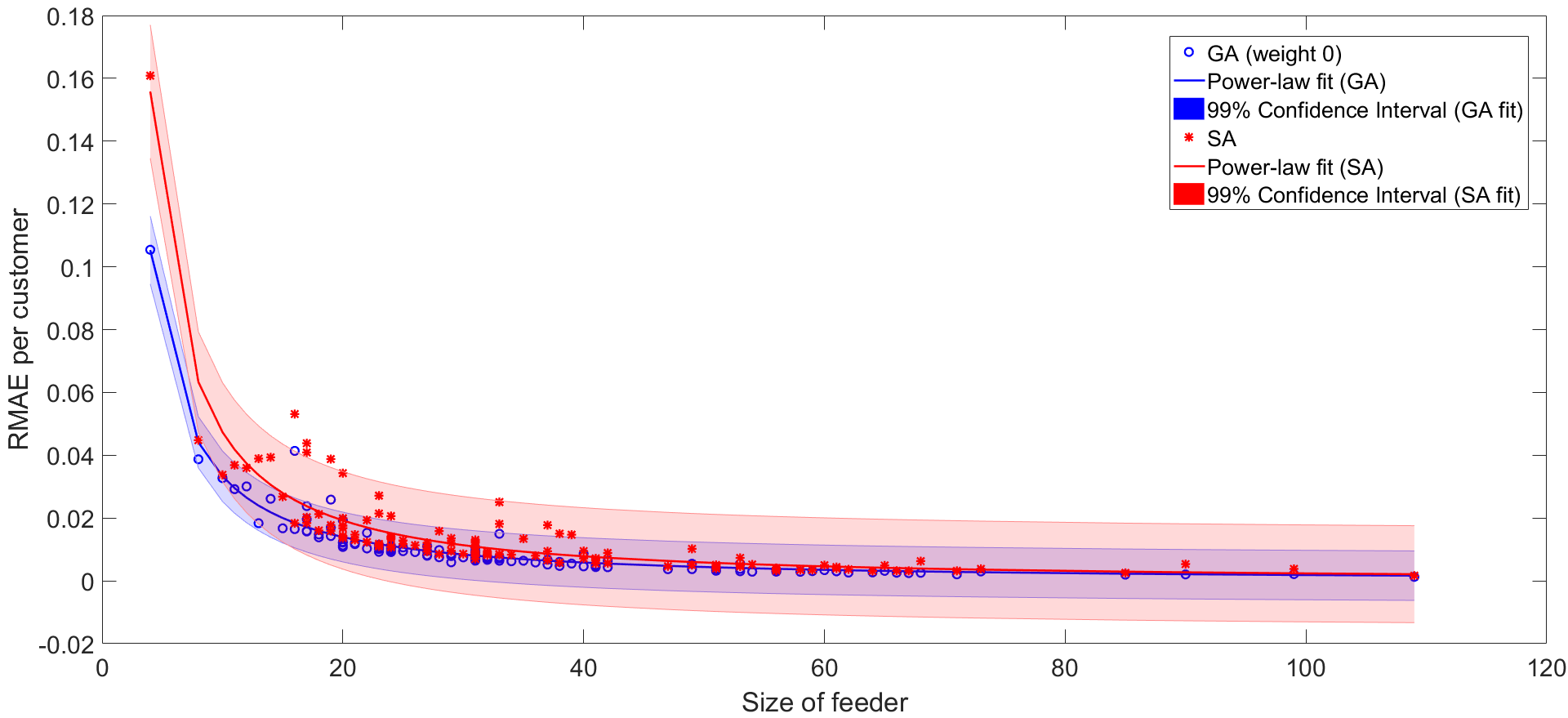}
	\caption{RMAE error per feeder as a function of the number of customers. Blue circles correspond to the GA with zero-weight and red stars correspond to the simple algorithm. 
	The solid blue and red lines correspond to power law fits on the GA and SA errors respectively, while the shaded bounds represent 99\% confidence intervals of the curves respectively.}
	\label{fig:error_distibution}
\end{center}
\end{figure}

An important question to a DNO is how the errors are distributed, and whether the errors are correlated to the other feeder characteristics. In Figure \ref{fig:error_distibution}, we plot the feeder's error score RMAE \eqref{eq:errormeasure1} divided by the size of the feeder as a function of the number of customers on the feeder, for two methods (i.e. 122 data-points per method), the GA with zero weight and the SA. The relationship of the errors with the number of customers resembles a power-law and hence we have included a power law fit of the form $a x^{-b}$, for positive $a, b$, on the GA and SA errors. Hence, using this curve we can thus estimate the size of buddying error, whatever the size of the feeder.

Several observations can be made from Figure \ref{fig:error_distibution}. First, we notice that the SA errors are larger than the GA ones. The SA errors are more variable for fixed numbers of customers, whereas the GA errors follow a power-law pattern more closely. The GA errors indicate that the buddying accuracy is greater for feeders with a higher number of customers. From a DNO perspective feeders with larger number of customers are more accurately modelled and thus potentially reduces the need for monitoring. On the other hand, feeders supplying a few customers have the lowest buddying accuracy, but they will have a low total demand leaving potentially more headroom, hence no monitoring is needed. It is feeders that have a significant total demand (usually supplying a few tens of customers), and relative high errors that might need further monitoring. Thus the power-law fit can guide DNOs about the expected modelling capabilities for a LV network of a certain size, and whether monitoring is required.

The power-law scaling behaviour is consistent to existing studies in the academic literature \cite{Sevlian2014}, where the authors studied the effect of aggregation on short term load forecasting. We leave a detailed comparison of our results to the conclusions of \cite{Sevlian2014} for future work.

\subsection{Comparison to Monte Carlo approach}
\label{sec:benchmark}
\begin{figure}
	\begin{center}
		\includegraphics[width = \columnwidth]{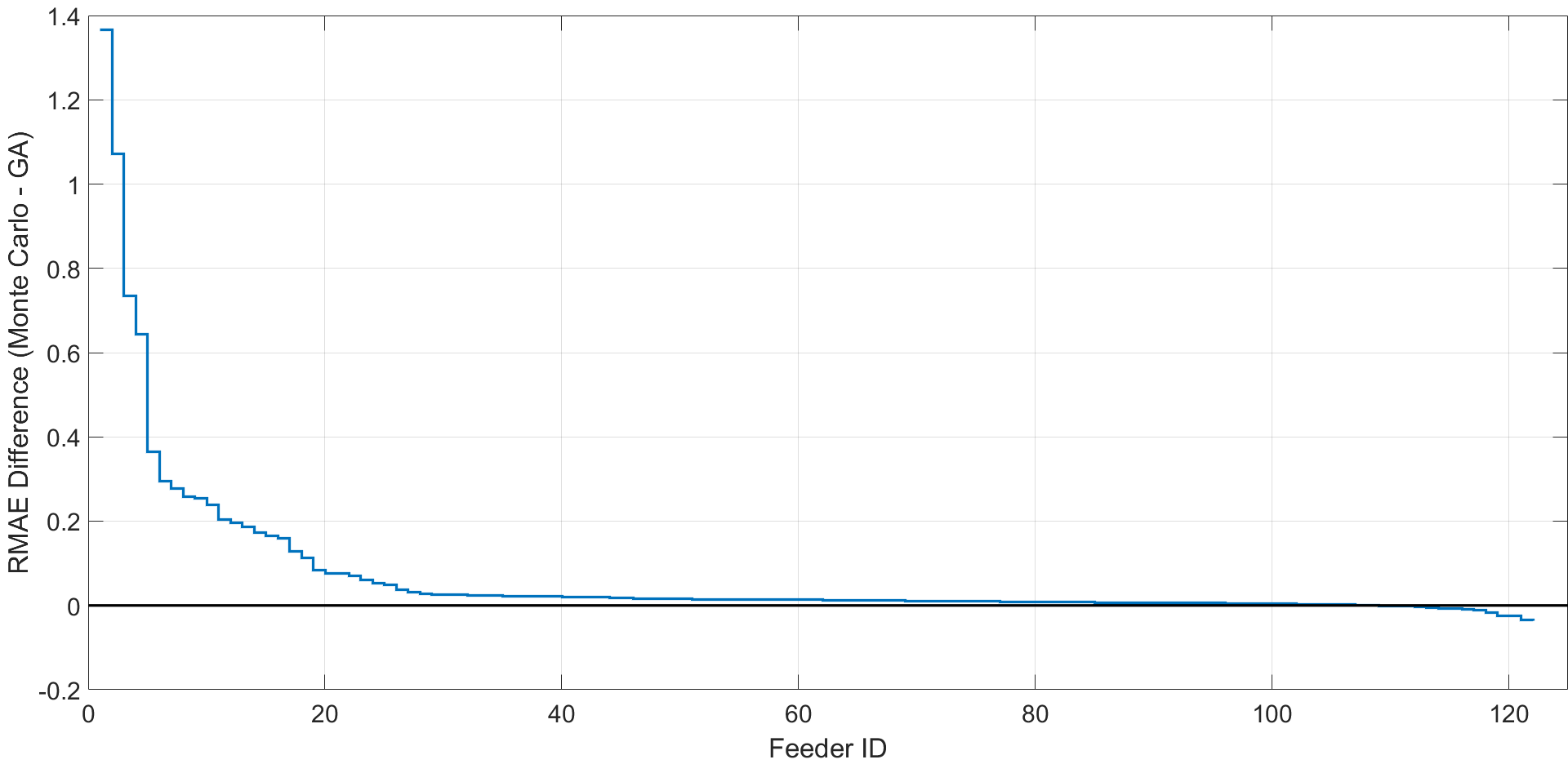}
		\caption{The difference of the RMAE errors between the GA and the best randomly sampled buddy for each of the 122 feeders ordered by the size of the difference. Positive values indicate that the Monte Carlo solution has larger errors than the GA.}
		\label{fig:benchmark_model_diff}
	\end{center}
\end{figure}

We compare the GA method for autumn, 8 weeks of training and zero weight to $1000$ randomly generated buddies, similar to a Monte Carlo simulation \cite{Neaimeh2015, Navarro2013}, so that the individuals are constrained to the same group of monitored customers. For each feeder, we choose the selection of buddies with the lowest RMAE as the solution of the Monte Carlo method.
The average RMAE of the Monte Carlo simulation is $0.322$ compared to $0.255$ of the GA with zero weight, see Figure \ref{fig:3Derrors}. In Figure \ref{fig:benchmark_model_diff}, we plot the RMAE (eq. \eqref{eq:errormeasure1}) difference between the GA buddying and the Monte Carlo solution. 

We observe that for 109 out of 122 feeders the GA method is more accurate than the Monte Carlo \cite{Neaimeh2015}. Particularly, in several cases the improvement by our model is substantial. For the remaining 13 feeders, ten feeders have their score in the best 2.5\% random samples, two in the best 5\% and the remaining one feeder has score in the best 30\%. For the vast majority of the feeders our method outperforms random sampling, showing the robustness of the method. As expected, the advantage of the genetic algorithm optimisation compared to a random search is that it ensures convergence to a global solution (or very close to global) in shorter execution times.

In general, $1000$ randomly sampled networks are less accurately modelled than a single run of the buddying algorithm. Thus the implication is that implementing a load flow analysis tool on the Monte Carlo technique is not only less efficient but also less accurate than the buddying approach.

\subsection{Comparison to buddying at phase-level}
\label{comparison_to_phaselevel}

Each feeder consists of three phases and customers on a feeder are connected to one of the phases. Buddying can also be performed at the phase-level considering only customers on a particular phase. However, at the phase-level, the exact phase allocation of the customers might not be known to a DNO. Therefore phases are allocated randomly introducing an extra degree of uncertainty. On the other hand, the genetic algorithm search is more likely to converge to a globally optimal solution when the number of customers is lower. This means that feeder-level buddying might not result in the globally optimal selection of monitored customers (especially for a feeder with a large number of customers). To study the effects of these two competing factors, i.e. the lack of phase information and the large number of customers, we also buddy and analyse at phase level. In particular, we buddy each phase of the 122 feeders separately and then aggregate the buddied profiles at the feeder level. We focus on autumn 2014 and train the algorithm for eight weeks for several weights. 

We find that simple algorithm's results remain the same when buddying either at the feeder or phase level, due to the fact that the simple algorithm makes no use of feeder or phase readings. For weights less than $0.8$, the buddying at feeder level outperforms the aggregate buddying at phase level. At zero weight the aggregate phase-level buddying achieves an average error score 0.263 which is an increase by 3.3\% compared to feeder-level buddying. 

\begin{figure}
	\begin{center}
		\includegraphics[width=\textwidth]{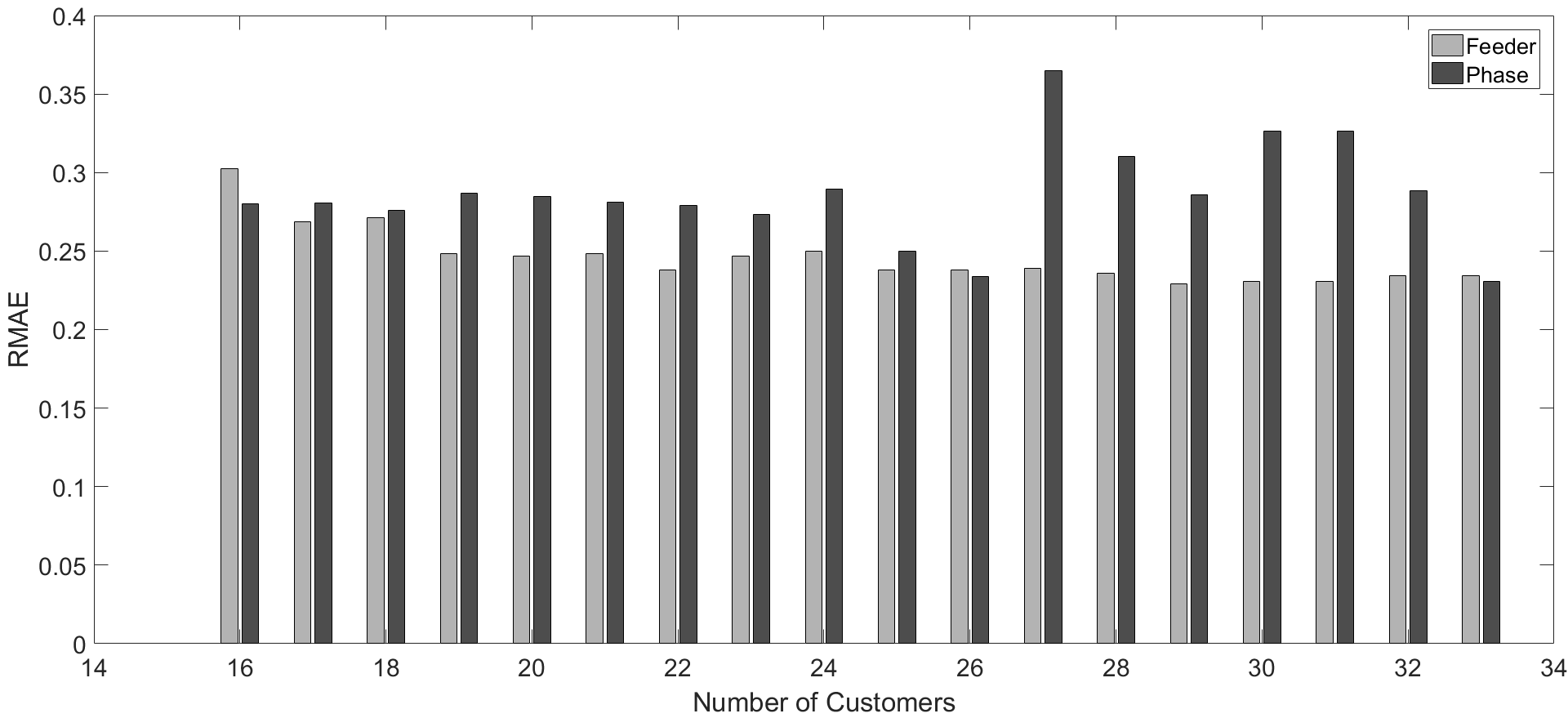}
		\caption{Average RMAE errors of feeders (light grey) and individual phases (dark grey) with population of $n$ and $n+1$ customers.}
		\label{fig:feeder_vs_phase_ncsts}
	\end{center}
\end{figure}

To study the effect of the lack of phase information, we compare the errors of individual phases to feeders with a similar number of customers on, see Figure \ref{fig:feeder_vs_phase_ncsts}. The training period is eight weeks in autumn 2014 with zero weight. We compare feeders to phases with a size ranging from $16$ to $36$ (there are only a few feeders with less than 16 customers on and similarly only a few phases with more than $36$ customers). To increase the sampling points and avoid cases where there exists only a single or no feeder (phase resp.) with a specific number of customers, we estimate the error as the average of all feeders (phases resp.) with population of $n$ and $n+1$ customers. Figure \ref{fig:feeder_vs_phase_ncsts} shows that the RMAE errors for feeders and phases with similar sizes are in general smaller for buddying at the feeder level.

 To conclude, the feeder-level buddying is insensitive to phase allocation and performs better than the phase-level buddying. This indicates that knowing the actual phase allocation likely improves the performance of the buddying. 

\subsection{Peak demand error}

\begin{figure}
	\begin{center}
		\includegraphics[width=\textwidth]{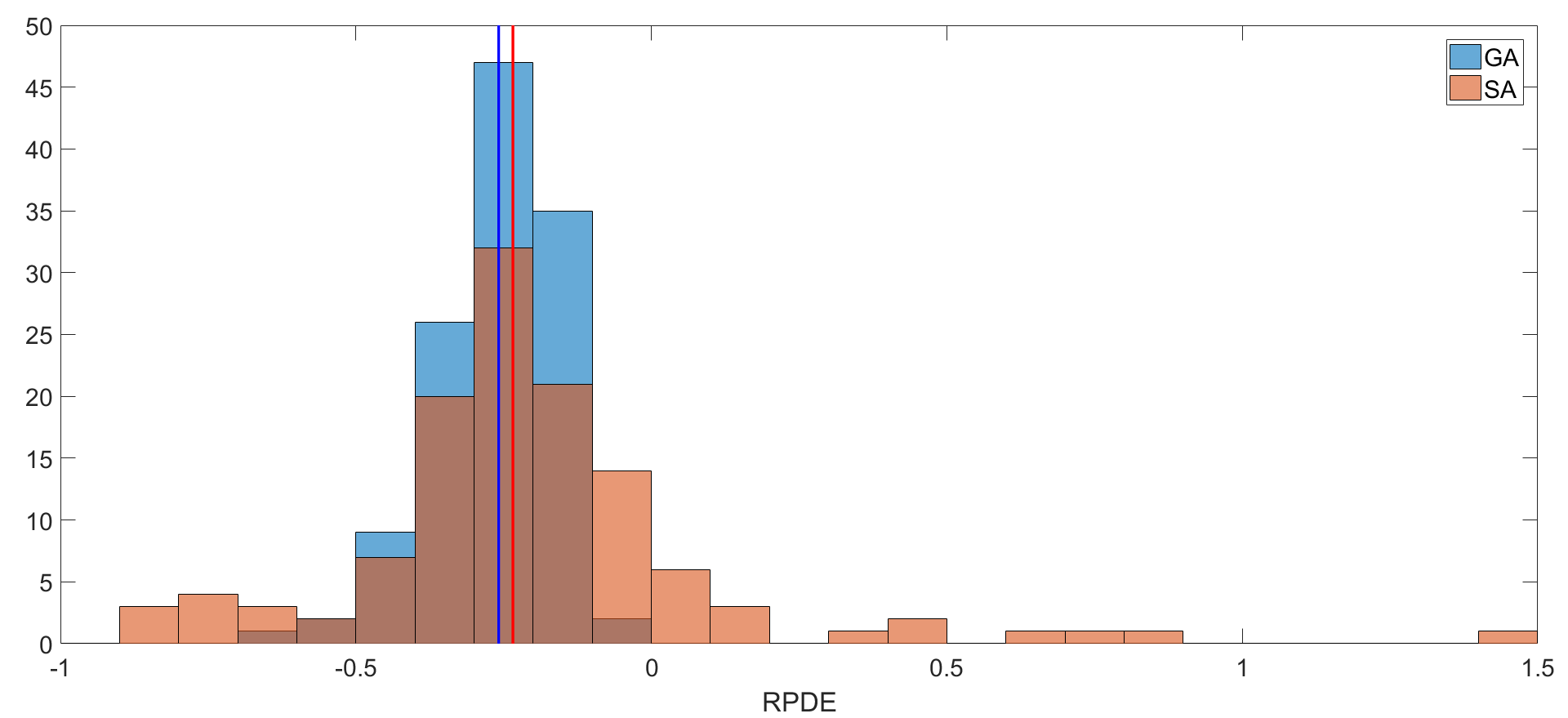}
		\caption{Distribution of relative peak demand errors (RPDE) for GA with zero weight and SA respectively. Vertical lines correspond to the median of GA (blue) and SA (red) RPDE respectively.}
		\label{fig:peak_error_boxplot}
	\end{center}
\end{figure}

A DNO is interested in the peak demands since they cause potential dangers to the network, such as breaching thermal constraints. We study the peaks using the error score \eqref{eq:errormeasure2}. In contrast to the time-series error \eqref{eq:errormeasure1}, the relative peak error can be negative (underestimate) or positive (overestimate). From a DNO perspective, the peaks must be estimated as closely as possible, therefore a good model is one with peak error close to zero, or a minimum absolute peak error. In Figure \ref{fig:peak_error_boxplot} we plot the distribution of feeder peak error for buddying trained for eight weeks in autumn 2014 with zero (i.e. GA) and unit weight (i.e. SA) respectively. 

We observe that the simple algorithm errors are more scattered and quite a few of them overestimate the peak. On the other hand the GA errors always underestimate the peak and their distribution is narrower. In both cases the median is negative and comparable. In this study, we trained our models so that the model matches the actual readings for the entire time-series. The model could have been trained to match the peak errors both in magnitude and in time of occurrence. An intermediate solution would be to consider a $p$-norm, $p>1$, in the first term of the fitness function \eqref{eq:CostFctn}; such cost functions penalise double peak errors more heavily. Alternatively, a simple correction can be applied to the estimate to ensure that the distribution of errors so that on average the peak error estimate is $0$. We leave improvements on the peak demand for future work.

\begin{figure}
 	\begin{center}
 		\includegraphics[width=\textwidth]{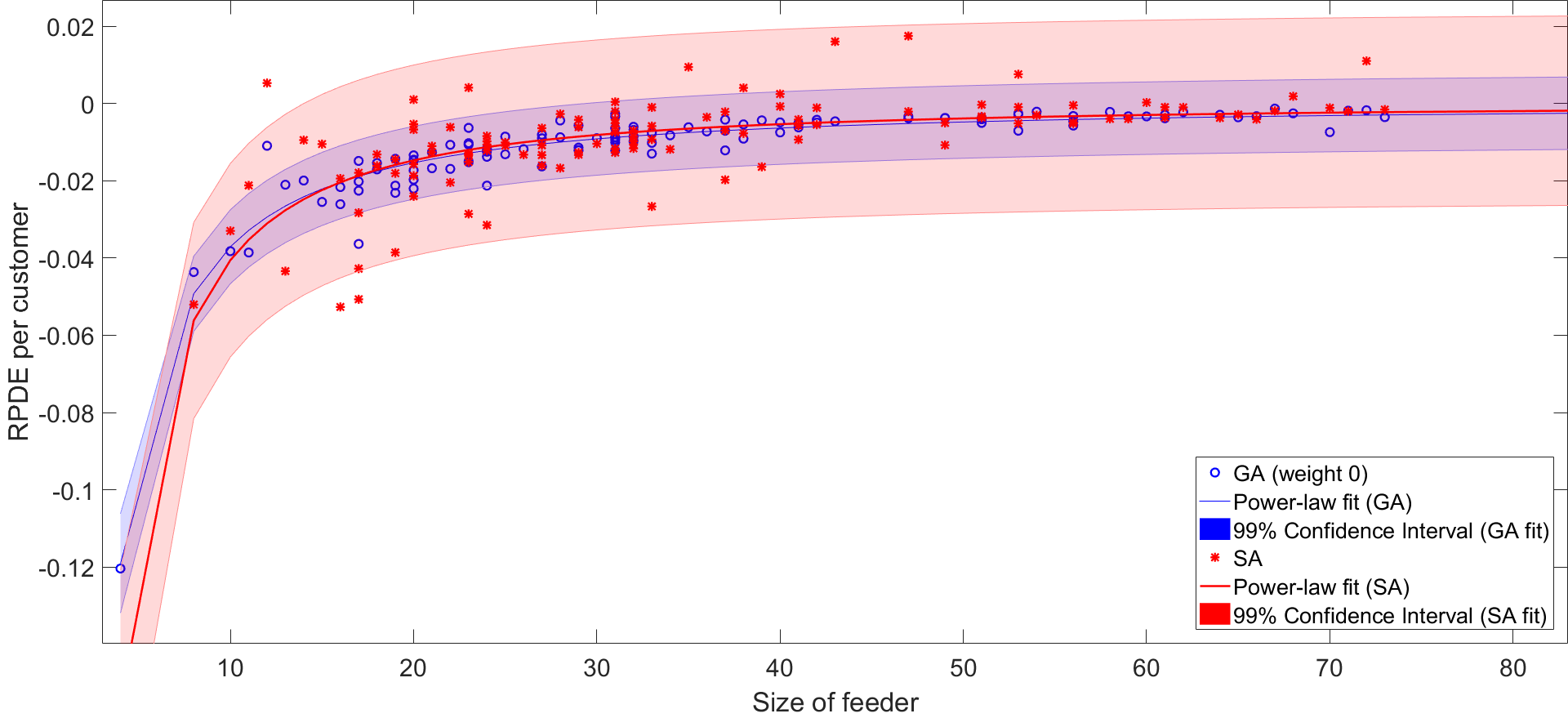}
 		\caption{Peak demand errors per customer for GA with zero weight and SA respectively against the feeder size. Blue circles correspond to the GA with zero-weight and red stars correspond to the simple algorithm. 
 			The solid blue and red lines correspond to power law fits on the GA and SA errors respectively, while the shaded bounds represent 99\% confidence intervals of the fits respectively.}
 		\label{fig:peak_error_vs_feedersize}
 	\end{center}
\end{figure}

In Figure \ref{fig:peak_error_vs_feedersize}, we also plot these errors per customer as a function of the feeder size. We notice a pattern where feeders with bigger sizes tend to have smaller absolute peak errors, i.e. closer to zero. Together with Figure \ref{fig:error_distibution} this plot can again help a DNO understand the modelling capabilities and limits of the buddying as well as help manage, plan and test the headroom of future LV networks,  for example, in creating new after diversity maximum demand estimates \cite{Frame2016}.

\section{Results on pseudo-feeders}
\label{sec:Pseudo_feeders}

A limitation of the study in the previous section is the small number of monitored customers on the network under consideration. Only a small percentage (1.2\%) of the customers on the feeders have been monitored and no feeder has more than three monitored customers connected. Hence, we cannot asses the accuracy of the buddying at the household level. For this reason, in addition to the analysis of buddying on real feeders, we also perform simulations on pseudo-feeders, i.e. feeders artificially and randomly populated with only monitored profiles, whose total demand is the aggregate of these profiles. To allow for comparison, we create a pseudo-feeder of the same size as every real feeder in our trial. Every pseudo-feeder is randomly populated with its original number of customers subject to group constraints. For example, a feeder with $25$ customers on, which consists of $10$ properties in group $0$, $7$ in group $1$, $8$ in group $2$ etc. will be assigned random monitored customers with the same group proportions.
      
We create two types of pseudo-feeders. In the first method, we use all the available monitored profiles to populate the pseudo-feeders. In the second method, we first split the set of monitored profiles into two subsets. One subset is used only for populating the pseudo-feeders, while the other subset is used as a new set of monitored profiles for buddying and assessing the estimation of individual household demand. 

\subsection{Pseudo-feeders type 1}
\label{sec:pseudo_type1}
In this method, the pseudo-feeders consist of customers which are also available in the set of monitored properties used for buddying. Using this method we test the accuracy of the algorithm to select the correct buddies. We focus on the autumn 2014 season and perform buddying for a number of training weeks, ranging from $1$ to $8$ and weight parameter ranging from $0$ to $1$ in $0.1$ increments. We calculate the percentage of properties that were correctly assigned by the algorithm to their identical profile from the monitored set. 

We find that the SA (i.e. unit weight buddying optimised to match the customers mean daily demand) is $100\%$ accurate. This is expected as the algorithm searches for the monitored profile with the closest daily usage.  
For weights smaller than one, the agreement percentage is also very high, exceeding $92\%$, in all cases, with only a few misassignments. This indicates that the algorithm operates successfully with very high accuracy rates. Second, the few mismatches are due to the nature of the genetic algorithm and indicate that a locally rather than globally optimal solution was found.

\subsection{Pseudo-feeders type 2}
\label{sec:pseudo_type2}

\begin{figure}
	\begin{center}
		\includegraphics[width=\textwidth]{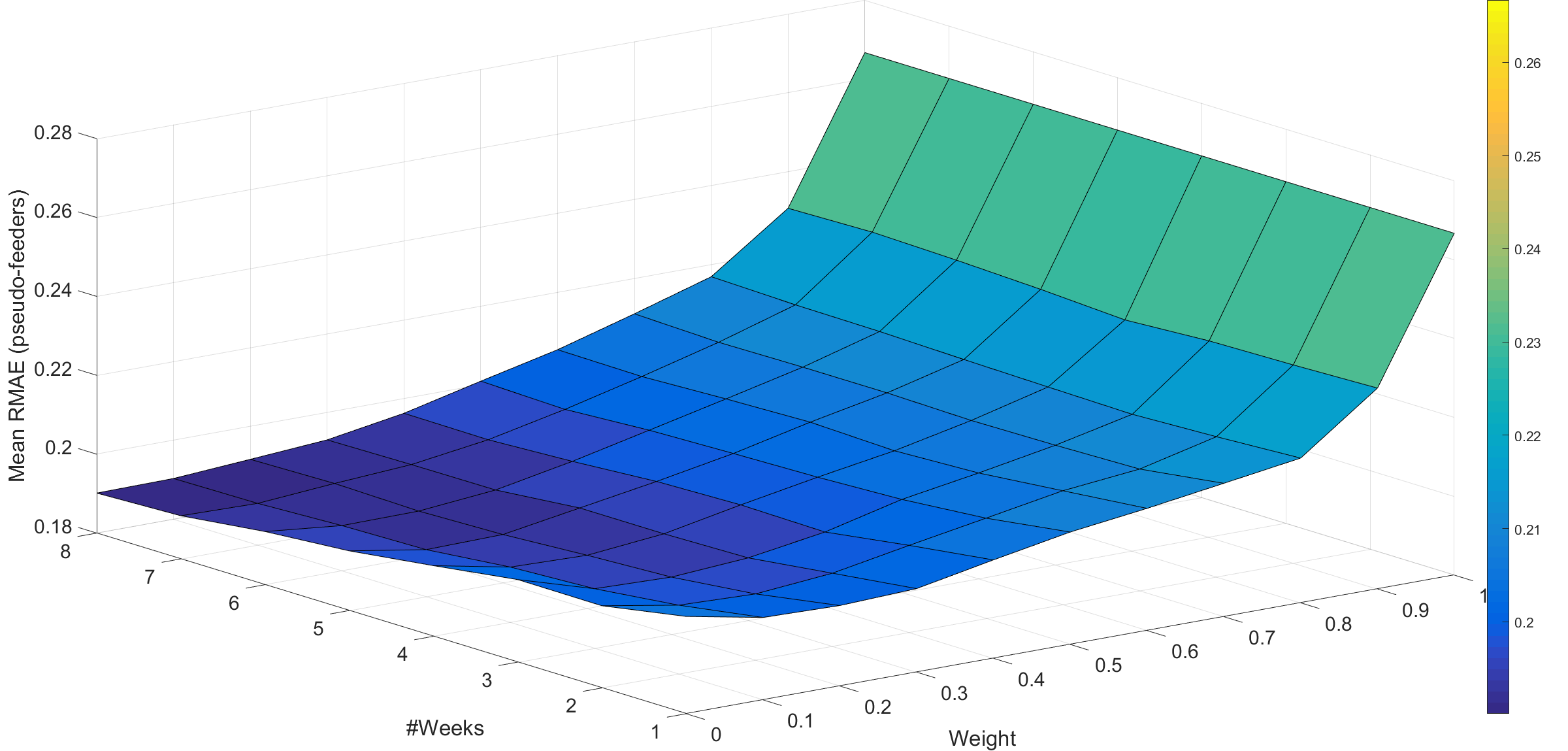}
		\caption{Average RMAE errors of all pseudo-feeders for season commencing on 29/09/2014 (autumn 2014). The z-axis is the average error score value, the x-axis, labelled as ``Weight'', corresponds to the weight in the fitness function of the genetic algorithm. The y-axis, labelled as ``\#Weeks'', is the number of weeks used in the training of buddying, each starting at the first date of the corresponding season.}
		\label{fig:pseudo_feeders_type2_feedererrors}
	\end{center}
\end{figure}

Using the second type of pseudo-feeders we aim to asses the accuracy of the individual profiles. For this reason we split the set of monitored profiles into two subsets. One subset is used to populate the feeders with customers and the other is used for buddying. This more closely resembles the buddying performed with actual feeders. Since we have limited monitored customers, and to ensure that the two subsets have properties with similar groups and mean daily demand, we order the profiles in each group by their mean daily demand in ascending order. We then split the profiles with even index into one group and those with odd index into the other. We trained the buddying for several weeks and weights in the period starting on 29/09/2014. Figure \ref{fig:pseudo_feeders_type2_feedererrors} shows how the average RMAE errors \eqref{eq:errormeasure1} of feeders vary as a function of weight and length of training (in weeks). This figure is consistent with Figure \ref{fig:3Derrors} for real data and similar conclusions can be drawn. First, the errors for pseudo-feeders are, on average, smaller than the errors of real data. This is likely because real feeders have other street furniture like street lights, traffic lights, etc. Second,  the GA with zero weight is 28.8\% more accurate than the SA, whereas it was 39\% more accurate when using real feeders. 

\begin{figure}
	\begin{center}
		\includegraphics[width=\textwidth]{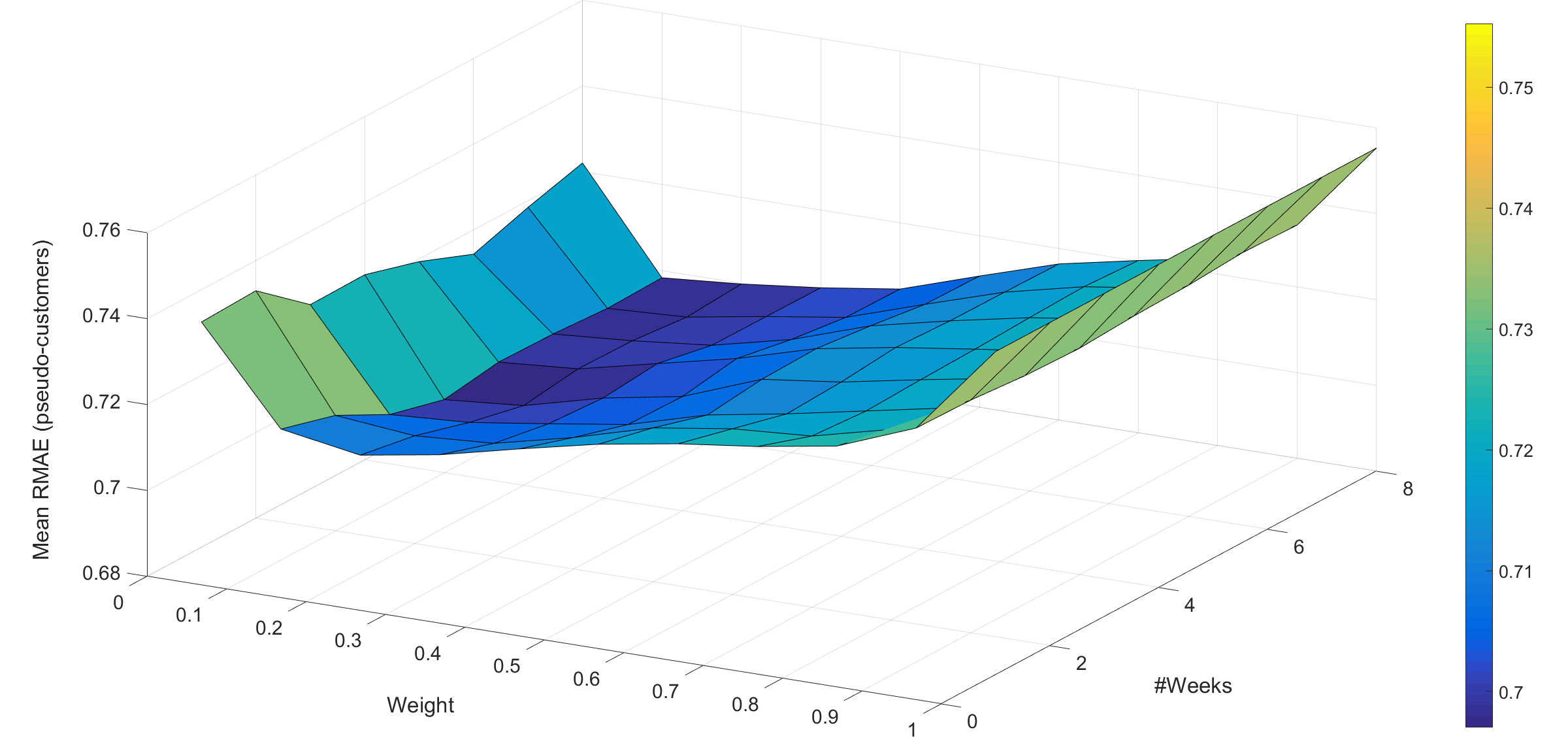}
		\caption{Average RMAE errors of individual properties on pseudo-feeders for season commencing on 29/09/2014 (autumn 2014). The z-axis is the average error score value, the x-axis, labelled as ``Weight'', corresponds to the weight in the fitness function of the genetic algorithm. The y-axis, labelled as ``\#Weeks'', is the number of weeks used in the training of buddying. This figure's axes have been rotated for better visualisation compared to the the previous three-dimensional figures.}
		\label{fig:pseudo_feeders_type2_mpanerrors}
	\end{center}
\end{figure}

In addition to the error at the feeder level, we are now able to estimate the error of the individual household profile assignments. Figure \ref{fig:pseudo_feeders_type2_mpanerrors} shows the average relative mean absolute error \eqref{eq:errormeasure1} of all buddied properties on the feeders under consideration. First we observe that the average normalised errors of individuals are much higher than the feeder level. This is expected because an individuals' profile is much more volatile than a feeder profile. However here we observe a slightly different pattern. Although the accuracy still increases for longer training periods, it is not monotonically increasing with decreasing weight. The best accuracy is achieved with eight-week training and $0.1$ weight. There are a number of interesting observations. Firstly, the optimal weight for modelling individual customers' profiles is not zero. This is an expected feature, because zero-weight buddying considers only the feeder profile and ignores the information on the individuals. Second, it is interesting that the optimal weight is not one (i.e. SA), in which case the algorithm would be optimised using only individuals information. Perhaps this is unsurprising since energy demand is only weakly correlated to mean daily demand \cite{McLoughlin2012}.
It seems that accurate individual profiles require heavy constraints on the feeder total (which contains the intra-day information), but also, some information on the customers' mean daily usage.

\section{Discussion} 
\label{sec:Discussion}

Individual household level data, from smart meter data, may not be readily available to DNOs and hence accurate models, which can simulate the demand of unmonitored customers in a power analysis tool of LV networks, are required.
In particular such analysis can help tailor DNO network plans so that operational management is less reactive and more anticipatory. In this paper we describe a ``buddying'' method for modelling residential customer demand behaviour which does not necessarily require a DNO to purchase large amounts of expensive smart meter~data. In particular the method ensures that we can assign realistic profiles to individuals rather than idealised profiles. Retaining the veracity of the demand profiles is important for any power analysis tool to understand the impact of spikey volatile demand at the LV level, especially when considering the impact of low carbon technologies on the network.

In this paper we have described two methods and a benchmark for modelling customer demand behaviour for unmonitored customers by buddying them to customers with monitors. In particular we have shown how both quarterly meter readings and substation monitoring can play an important role in creating an accurate buddy over a years worth of data using eight weeks of substation monitoring data. Buddying can accurately estimate daily demand on phases of LV substations with as few as ten customers. The GA buddying also more accurately estimates the substation peak demand compared to the other methods tested. We also found important relationships between the errors in the buddies and the numbers of customers on the feeder. This is especially important for a DNO who could use such relationships in planning and managing the networks, identifying where modelling can be appropriate and targeting feeders with monitoring. However, a drawback to the method is that connectivity information is important for ensuring accurate buddies. This is illustrated in Section \ref{comparison_to_phaselevel} where phase level buddying with unknown connectivity is compared to feeder level with known connectivity. Additionally, the current version of buddying does not address a series of questions, such as the timing of the peak, underestimating the peak demand, uncertainty bounds, etc. These are all very important questions and will be the subject of future work.

There is still much further work to be investigated. In particular the buddying itself is a baseline for further modelling. The profiles can be augmented with low carbon technologies profiles to consider the impact of future scenarios. Further, the buddying itself allows forecasting to be implemented on unmonitored networks which could be utilised in the control of smart control of storage devices \cite{Rowe2014}. However this incorporates two levels of uncertainty via the buddying and the forecasting. Further research would be required in order to optimise such a process.

\section*{Acknowledgements}
We thank Scottish and Southern Electricity Networks (SSEN) for their support via the New Thames Valley Vision Project (SSET203 New Thames Valley Vision), funded through the Low Carbon Network Fund. We would also like to thank Helen Waller from the SSEN for her assistance in data collection and distribution.

\appendix

\section{Property grouping}
\label{sec:Grouping}

The genetic algorithm searches for the best combination of profiles from a set of monitored profiles. An average feeder consists of 40 properties which must select one of the 242 monitored profiles. Hence the possibilities are vast and the algorithm might not converge or it might converge to a local solution away from the global one. 
For this reason, the search space in the genetic algorithm is restricted to properties that share some characteristics, such as their profile class and the council tax band,\footnote{Council Tax is a local taxation system used in Great Britain on domestic properties. Each property is assigned one of eight bands (A to H) based on property value.} which ensures that the algorithm converges faster to a solution. 

We split properties into groups according to their profile class (given the large difference in the energy behavioural characteristics), council tax band and the presence of photovoltaic equipment. The current groupings are given in Table \ref{tbl:groups}.

\begin{table}
	\begin{center}
		\begin{tabular}{ | c | c | c | c | c | }
			\hline
			Group & Profile Class & Council Tax Band & Photovoltaic (Y/N) \\ \hline \hline
			0 & 1 & A, B, C 	& N	 \\ \hline 
			1 & 1 & D 			& N  \\ \hline  			
			2 & 1 & E			& N  \\ \hline 
			3 & 1 & F, G, H		& N  \\ \hline 
			4 & 2 & Any			& N  \\ \hline 
			5 & 2 & Any 		& Y  \\ \hline 
			6 & 1 & Any			& Y  \\ \hline 
		\end{tabular}
		\caption{Group IDs and their characteristics.}
		\label{tbl:groups}
	\end{center}
\end{table}

It is important to highlight that in regions, like in the USA, where profile class and council tax band information do not exist, one can use different groupings based on other characteristics, e.g. MOSAIC classifications. Both the simple and the genetic algorithms require a group ID, but the details of the groupings depend on the application.

\section*{References}

\bibliographystyle{elsarticle-num}
\bibliography{mybibfile}

\end{document}